\documentclass[journal,twoside, fleqn]{IEEEtran}
\pdfoutput=1 
\input{header.cls}

\begin{document}   

\title{The Sharing Economy for the Smart Grid$^\pi$}
%

\author{%
Dileep~Kalathil$^a$,~%
Chenye~Wu$^a$,~%
Kameshwar Poolla$^b$,~%
Pravin~Varaiya$^a$%
\thanks{$^\pi$Supported in part by  CERTS under sub-award 09-206, NSF under Grants ECCS-0925337, 1129061, CNS-1239178.}
\thanks{$^a$ Electrical Engineering \& Computer Science, Univ. of California, Berkeley, USA.}%
\thanks{$^b$Corresponding author, Electrical Engineering \& Computer Science, Univ. of California, Berkeley, USA, poolla@berkeley.edu}
}


\maketitle

\begin{abstract}

The sharing economy has upset market for housing and transportation services. Homeowners can rent out their property 
when they are away on vacation, car owners can offer ridesharing services. 
These sharing economy business models are based on monetizing under-utilized infrastructure. They are enabled by peer-to-peer
platforms that match eager sellers with willing buyers.

Are there compelling sharing economy opportunities in the electricity sector? What products or services can be
shared in tomorrow's Smart Grid? 
We begin by exploring sharing economy opportunities in the electricity sector, and discuss regulatory and
technical obstacles to these opportunities. We then study the specific problem of a collection of firms sharing their electricity storage.
We characterize equilibrium prices for shared storage in a spot market.
We formulate storage investment decisions of the firms as a non-convex non-cooperative game. We show that under a mild alignment condition,
a Nash equilibrium exists, it is unique, and it supports the social welfare.
We discuss technology platforms necessary for the physical exchange of power, and market platforms necessary to trade electricity storage.
We close with synthetic examples to illustrate our ideas.
\end{abstract}


\section{Sharing in the Electricity Sector}

The sharing economy. It is all the rage. Going on vacation? Rent out your home for extra income! Not
using your car. Rent it out for extra income! Companies such as AirBnB, VRBO, Lyft, and Uber are
disrupting certain business sectors \cite{sharing_the_economist}. Their innovative business models are based on resource sharing that leverage underutilized infrastructure. And much of our infrastructure is indeed underused. Some
90\% of cars are simply parked 90\% of the time. Investors have rewarded companies in this new sharing
economy model. For example, privately held Uber reached a valuation of \$60 B  in December 2015. 
more valuable faster than both Google and Facebook.

\subsection{Sharing Opportunities}
 
To date, sharing economy successes in the grid have been confined
to crowd-funding for capital projects \cite{sharing_greentech}.  What other products or services could be shared in tomorrow's grid?
We can imagine three possibilities. Many others surely exist.

(a) {\em Sharing excess generation from rooftop PV} \\
Surplus residential PV generation is sold today through net-metering programs.
Here, utilities purchase excess residential generation at a fixed price $\pi_{nm}$. Usually $\pi_{nm}$ is the retail price 
and there is an annual cap, so households cannot be net energy producers over the course of a year.  Utilities are
mandated to offer net-metering, but do so reluctantly and view such programs with hostility as it threatens their profitability and business model \cite{WashPost}.
Net-metering is not, strictly-speaking, sharing.
True resource sharing would pool excess PV generation and trade this over a spot market. Utilities could be paid a toll for
access to their distribution infrastructure.

(b)  {\em Sharing flexible demand recruited by a utility} \\
Many consumers have flexibility in their electricity consumption patterns.
Some consumers can defer charging their electric vehicles, or modulate use of their AC systems.
Utilities are recognizing and monetizing the value of this demand flexibility. Excess recruited demand flexibility could be shared
and used at other buses where generation is expensive. Trading this shared resource requires infrastructure to
coordinate the physical power transactions and support financial transactions.

(c) {\em Sharing unused capacity in installed electricity storage} \\
Firms faced with time-of-use (ToU) pricing might invest in storage if it is sufficiently cheap.  These firms could displace
some of their peak period consumption by charging their storage during off-peak periods when electricity is cheap, and discharging it during peak periods when it is dear.
On days when their peak period consumption happens to be low, these firms may have unused storage capacity.
This could be sold to other firms. This sharing economy opportunity is the focus of this paper.

\subsection{Challenges to Electricity Sharing Business Models}

A principal difficulty with sharing economy business models for electricity is
in tracing power flow point-to-point \cite{bialek1996tracing}. Electricity injected at various nodes and extracted at others flows according to Kirchoff's Laws, and we cannot assert that a KWh of electricity was sold by party $\alpha$ to party $\beta$.  As a result, supporting peer-to-peer shared electricity services requires coordination in the hardware that transfers power \cite{he2008architecture}. An alternative is to devise pooled markets which is possible because electricity is an undifferentiated good.
Regulatory and policy obstacles may impede wider adoption of sharing \cite{cannon2014uber}. The early adopters will use behind-the-meter opportunities such as in industrial parks or campuses, where sharing can be conducted privately without utility interference.

The successes of AirBnB or Uber are, to a large extent, resulted by  their peer-to-peer  sharing platform.
This brings together willing sellers and eager buyers and enables to settle on mutually beneficial transactions.
In the electricity sector, the challenges are to develop (a) software platforms that support trading, and (b) hardware platforms that realize the associated physical transfers of electricity.
These must be scalable, support security, and accommodate various market designs.

%
%

\subsection{Our Research Contributions}

We study the specific problem of a collection of firms sharing their electricity storage. The principal contributions of this paper are:

\setlength{\leftmargini}{0.1in}
\bdashlist
\item {\sl Spot Market for Sharing:} We formulate storage sharing in a spot market, and characterize the random clearing prices in this market.
\item {\sl Investment Decisions:} We explicitly determine the optimal investment decisions of a firm under sharing as the solution of a two-stage optimization problem.  
\item {\sl Equilibrium Characterization:} We formulate the investment decisions of a collection of firms as a Storage Sharing Game. This is a nonconvex game.
Under a mild alignment condition, we show that this game admits a Nash Equilibrium. We further show that if a Nash equilibrum exists, 
it is unique. We  explicitly characterize  the optimal investment decisions at this Nash equilibrium.
\item {\sl Social Welfare:} We show that these optimal investment decisions form a Nash equilibrium which supports the social welfare. 
\item {\sl Neutrality of Aggregator:} We show that the aggregator serves to inform firms of their optimal storage investments while protecting their private information. 
\item {\sl Coalitional Stability}: We prove that at this Nash equilibrium, no firm or subset of firms is better off defecting to form their own coalition. 
\item {\sl Implementation:} Under sequential decision making, we propose a natural payment mechanism for new firms to join the sharing coalition and realize this Nash Equilibrium.
\end{list}

\subsection{Related Work}

There is recent literature on estimating the arbitrage value and welfare effects of storage in electricity markets. 
Graves \emph{et al.} study the value
of storage arbitrage in deregulated markets \cite{graves1999opportunities}.  
Sioshansi \emph{et al.} explore the role of storage in wholesale electricity markets \cite{sioshansi2009estimating}. 
Bradbury \emph{et al.} examine the economic viability of the storage systems through price arbitrage  in \cite{bradbury2014economic}.  
Van de Ven \emph{et al.} propose an optimal control framework for end-user energy storage devices in \cite{van2013optimal}.  
Zheng \emph{et al.} introduce agent-based models to explore tariff arbitrage opportunities for residential storage systems  \cite{zheng2014agent}. 
Bitar \emph{et al.} characterize the marginal value of co-located storage in firming intermittent wind power \cite{Bitar2016}.
While these previous works illuminate the economic value of storage to an individual, to the best of our knowledge, the analysis of \emph{shared electricity services} has not been addressed in the literature.

\section{Problem Formulation}


For a random variable $X$, its expectation is written $\EXP{X}$, the probability of some event $\Acal$ is written $\Prob{\Acal}$, and we define
$x^+ = \max\{x,0\}$.

\begin{figure}[h!]
\captionsetup{justification=centering}
\centering
\includegraphics[scale=0.55]{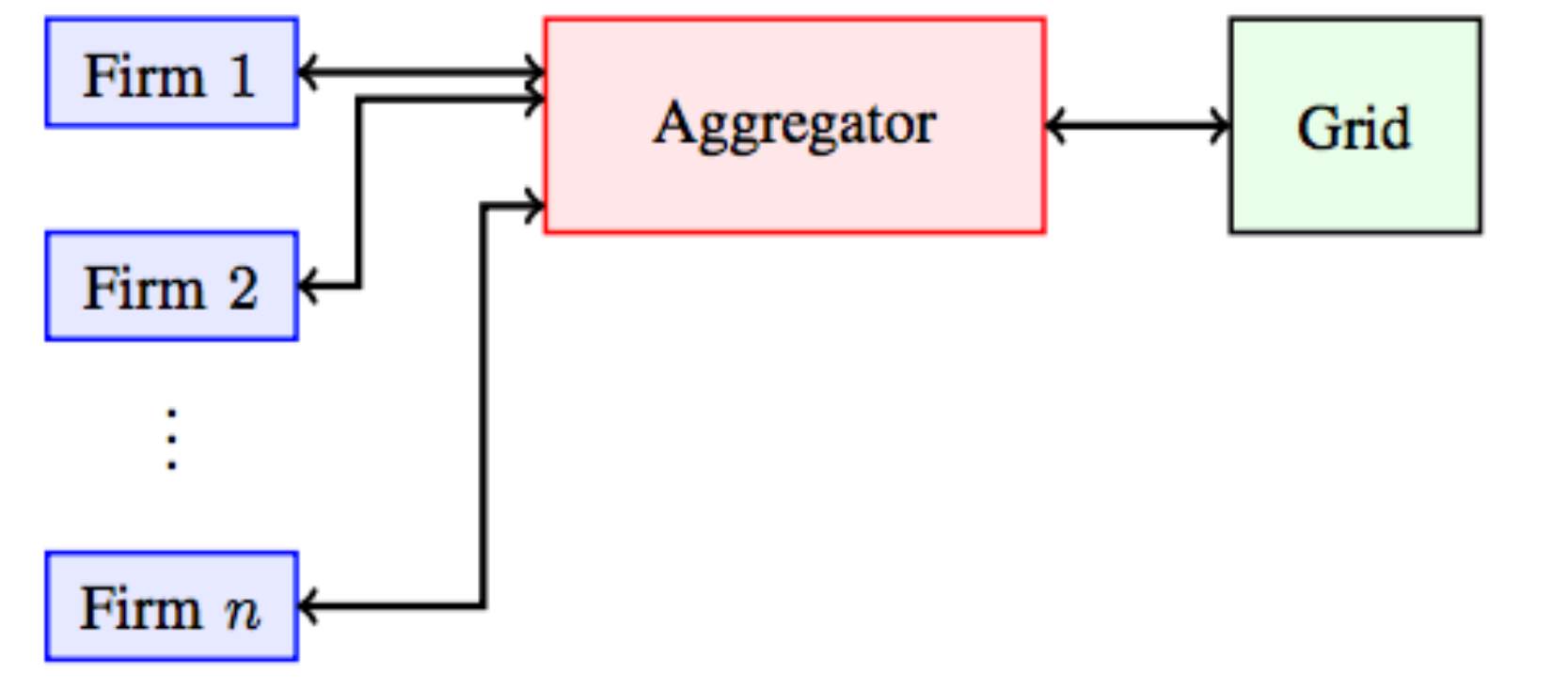}
\caption{Agents and interactions.} \label{fig:agents}
\end{figure}

Consider a collection of $n$ firms indexed by $k$ that use electricity. An aggregator interfaces between these firms and the utility.
The aggregator itself does not consume electricity. It purchases the collective electricity needed by the firms from the utility,
and resells this to the firms at cost. Firms can trade electricity with each other,
or purchase electricity from the utility through the intermediary aggregator. The physical delivery of electricity for these transactions are
conducted over a private distribution system within the aggregators purview. Prices imposed by the utility are passed through to the firms.
The aggregator does not have the opportunity to sell excess electricity back to the utility (no net metering).
The situation we consider is illustrated in Figure \ref{fig:agents}.

%

\begin{remark}
Examples of the situation we consider include firms in an industrial park, office buildings on a campus, or homes in a residential complex. 
The aggregator might be the owner of the industrial park, the university campus, or the housing complex community. 
There is a single point of  coupling or metered connection to the utility. 
Exchanges of energy between firms, buildings, or homes are {\sl behind-the-meter} private
transactions outside the regulatory jurisdiction of the utility.  The distribution grid serving these firms, buildings, or homes is private. It could be
communally owned or provided by the aggregator for a fee. If this fee is a fixed connection charge, 
our results are unaffected. Analysis of sharing when the distribution system charge is proportional to use is substantially more complex
and falls outside the scope of this paper. We ignore capacity constraints in this private distribution grid, and our results are agnostic to
its topology. We also ignore line losses in the distribution system. \hfill $\Box$
\end{remark}

\begin{figure}[h!]
\captionsetup{justification=centering}
\centering
\includegraphics[scale=0.4]{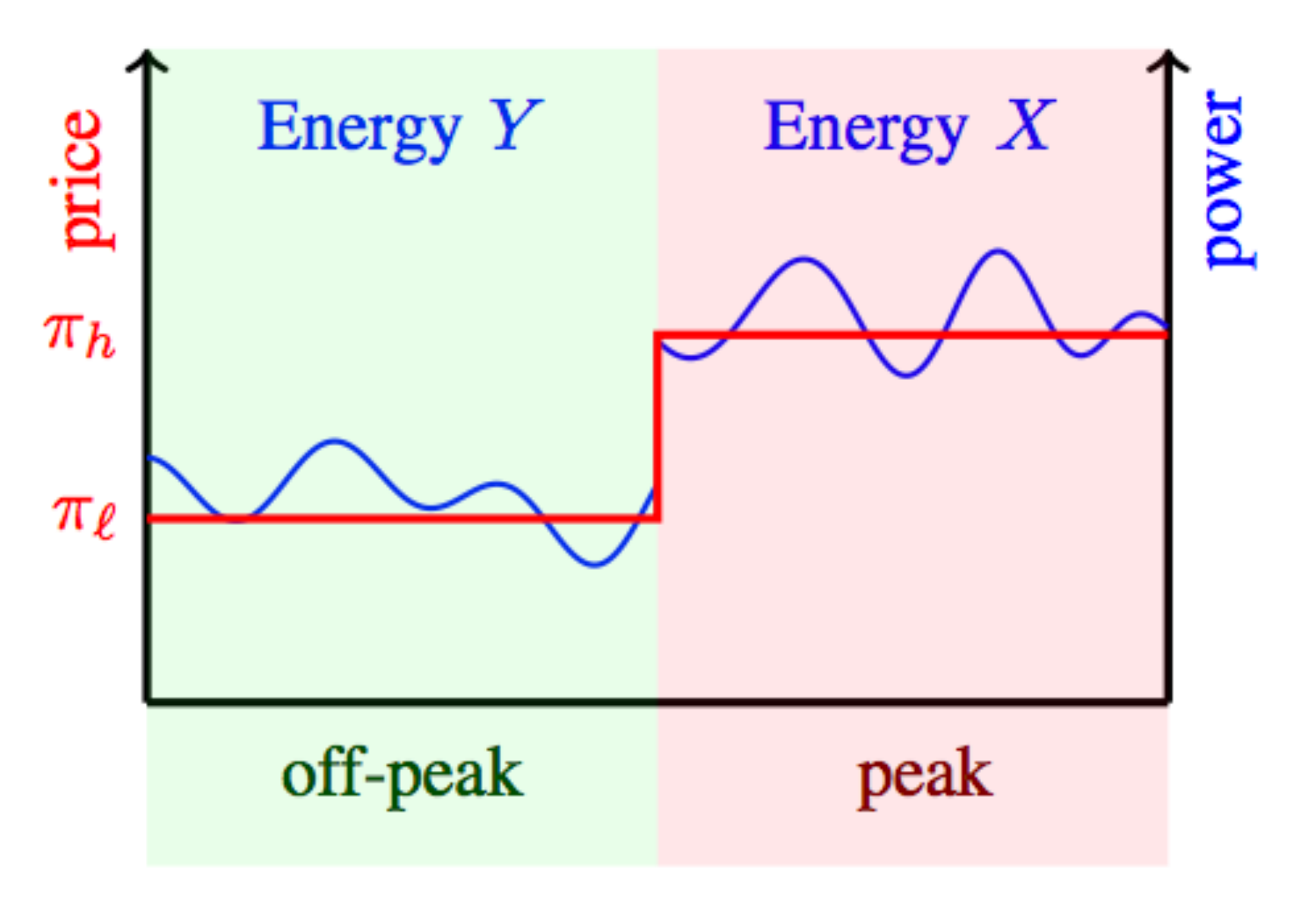}
\caption{Consumption and Time-of-Use Pricing.} \label{fig:tou}
\end{figure} 

Each day is divided into two fixed, contiguous periods -- peak and off-peak.
The firms face common time-of-use (ToU) prices.  During peak hours, they face a price $\pi_h$, while during off-peak hours,
they face a lower price $\pi_\ell$.
These prices are fixed and known. Our formulation considers the simplest situation of two-period ToU pricing.
Weekday only ToU pricing is easily handled.

The consumption of firm $k$ during peak and off-peak hours on day $t$ are the random processes $X_k(t)$ and $Y_k(t)$ respectively.
We model  $X_k$ as an independent identically distributed random sequence.
Let $F_{k}(\cdot)$ be the cumulative distribution function of $X_k(t)$ for any day $t$. 
Let $f_k(\cdot)$ be the probability density function of $X_k(t)$ for any day $t$.
Empirical distributions of $X_k$ may be estimated from historical data using standard methods \cite{van2000asymptotic}. 
Let 
\beq X_c = \sum_k X_k \eeq
be the collective peak-period consumption of all the firms. The cumulative distribution function and probability density function for $X_c$ are
$F_c(\cdot)$ and $f_c(\cdot)$ respectively.
Consumption and pricing are illustrated in Figure \ref{fig:tou}.

The  off-peak consumption $Y_k$ is not material to our results. This is because it is serviced at  $\pi_\ell$ which is the lowest price 
at which electricity is available. The use of storage cannot reduce this expense. We therefore disregard $Y_k$ in the remainder of this paper.

\begin{remark}
We can just as easily treat the case where $X_k$ is
not stationary. In this case, all stated results hold with $F_k$ being replaced by the average of the cumulative distribution functions over the $T$-day
lifetime of electricity storage:
\[ F_k(\cdot) \rightarrow \frac{1}{T} \sum_t F_{X_k(t)}(\cdot) \hfill \Box \]
\end{remark}

If storage is sufficiently cheap, firms will invest in storage to arbitrage ToU pricing.
Let $\pi_s$ be the daily capital cost of storage amortized over its lifespan.
Define 
\begin{eqnarray}
\text{arbitrage price} &  \pi_\delta = \pi_h - \pi_\ell  > 0 \\[0.05in]
\text{arbitrage constant} & \dst{\gamma = \frac{\pi_\delta - \pi_s}{\pi_\delta}}
\end{eqnarray}
For storage to offer a viable arbitrage opportunity we clearly require 
\beq \pi_s < \pi_\delta \eeq
In this case,  $0 < \gamma < 1$.
With this assumption it is profitable for firms to invest in storage.
They charge their storage during off-peak hours when electricity is cheap, and discharge it during peak hours when it is dear.
Note that the energy that is held in storage is always acquired at price $\pi_\ell$/MWh. Let $C_k$ be the storage investment of firm $k$, and let
\beq C_c = \sum_k C_k \eeq
be the collective storage investment of all the firms.

\begin{remark}
Electricity storage is expensive.  The amortized cost of Tesla's Powerwall Lithium-ion battery is around 25\textcent/KWh per day \cite{storage_price}
(assuming one charge-discharge cycle per day over its 5 year lifetime). 
At current storage prices, ToU pricing rarely offers arbitrage opportunities.
An exception is the three-period PG\&E A6 program \cite{tou_price} under which 
the electricity prices per KWh are 54\textcent\ for peak hours (12:00pm to 6:00pm),
25\textcent\ for part-peak hours (8:30am to 12:00pm, and 6:00pm to 9:30pm), and 18\textcent\ for off peak hours (rest of the day).  Storage prices are projected to decrease by 30\% by 2020 \cite{storage_price}. Our results offer a framework for the analysis of sharing in this future of cheap electricity storage prices with lucrative sharing opportunities. \hfill $\Box$
\end{remark}


We make the following assumptions. 
\setlength{\leftmargini}{0.1in}
\begin{list}{A.\arabic{l1}}{\usecounter{l1}} 
\item Arbitrage opportunity exists:  $\pi_\delta > \pi_s$.
\item Firms are price takers: the total storage investment is modest and does not influence the ToU pricing offered by the utility.
\item Inelastic demand: the statistics of the demand $X_k, Y_k$ for firm $k$ are not affected by savings from ToU arbitrage.
\item Statistical assumptions: $f_k(\cdot)$ is continuously differentiable and
 $f_c(x) > 0$ for $x \geq 0$.
\item Electricity storage is ideal: it is lossless, and perfectly efficient in charging/discharging.
\item Storage investments by the firms are made simultaneously.
\end{list}
Assumptions A.1 through A.3 are essential. Assumption A.4 is needed only to simplify our exposition. We will dispense with  A.5 and A.6 in Sections \ref{sec:gen} and \ref{sec:club} respectively.

\section{Main Results: No Sharing}

\subsection{Optimal Investment Decisions}

We first consider a single firm which elects to invest in storage capacity $C$. 
Let $X,Y$ be the random peak and off-peak consumption of this firm.
The firm will choose to service $X$ first using its cheaper stored energy, and will purchase the deficit 
$(X -C)^+$ at the peak period price $\pi_h$. During the off-peak period, it will service its demand $Y$ and recharge its storage 
at the lower price $\pi_\ell$. The firm will elect to completely recharge the storage as we have assumed the storage is ideal, and
holding costs are zero. The daily expected cost of the firm is therefore
\beq \label{eq:costa}
 J(C) = \underbrace{\pi_s C}_\text{\footnotesize cap ex} + \underbrace{\pi_h \EXP{(X - C)^+}}_\text{\footnotesize buy deficit} 
 + \  \underbrace{\pi_\ell \EXP{\min\{C, X\}}}_\text{\footnotesize recharge storage}
\eeq
This firm will choose to invest in storage capacity 
\[ C^* = \arg \min_C J(C) \] 

\begin{theorem} \label{thm:1}
The optimal decision of a standalone firm under no sharing is to purchase $C^o$ units of storage where
\beq F(C^o) = \frac{\pi_\delta - \pi_s}{\pi_\delta} = \gamma \eeq
The resulting optimal cost is
\beq \label{eq:optcosta} J^o = J(C^o) = \pi_\ell \EXP{X}  +  \pi_s \EXP{X \mid X \geq C^o}  \eeq
\end{theorem}

\begin{figure}[h!]
\captionsetup{justification=centering}
\centering
\includegraphics[scale=0.4]{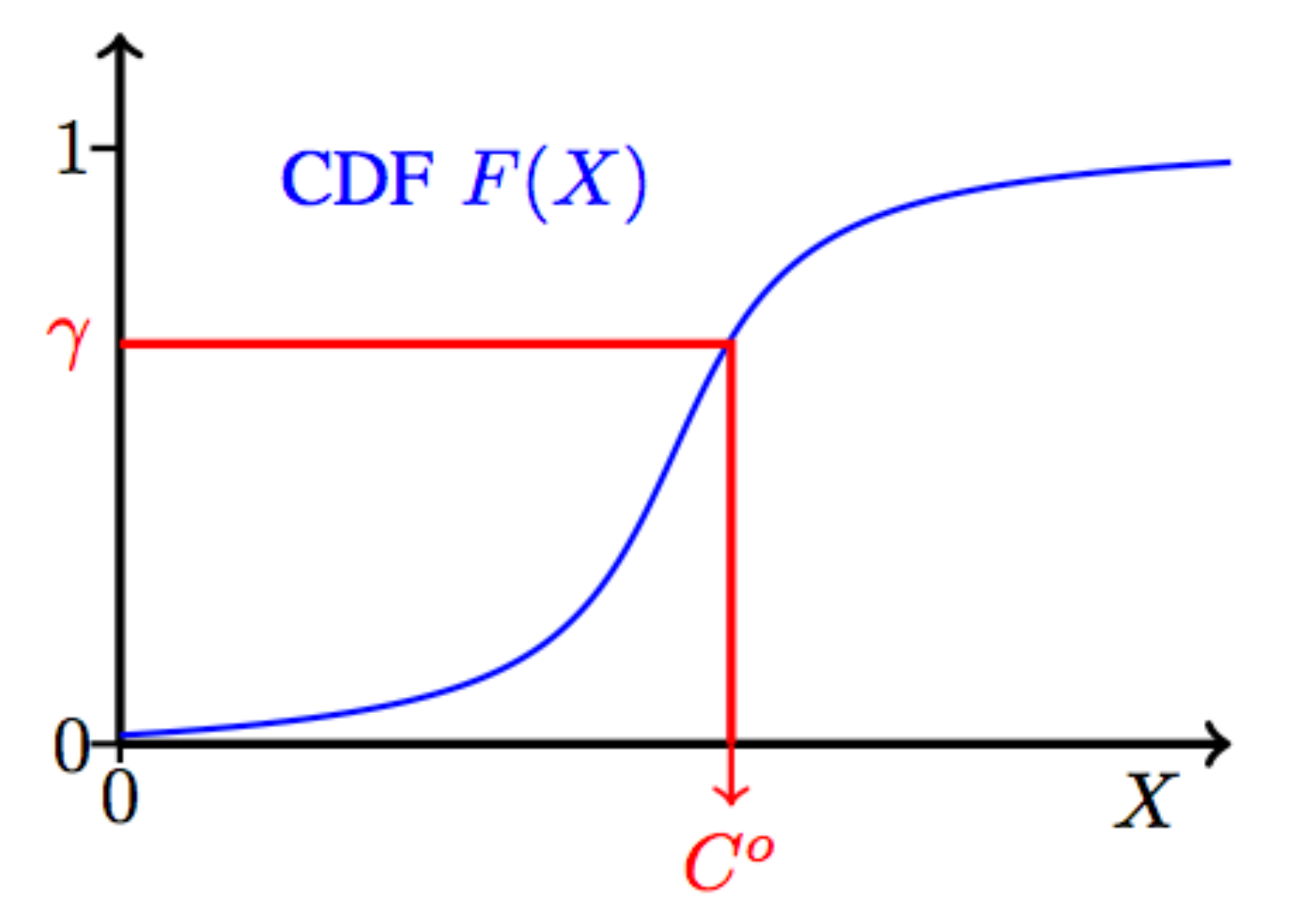}
\caption{Optimal Storage $C^o$ for a standalone firm.} \label{fig:copt}
\end{figure}

\begin{remark} The result above is illustrated in Figure \ref{fig:copt}. 
It is easy to show that the optimal storage investment $C^o$ is monotone decreasing
in the amortized storage price $\pi_s$, and monotone increasing in the arbitrage price $\pi_\delta$. \hfill $\Box$
\end{remark}

%

\begin{figure}[h!]
\captionsetup{justification=centering}
\centering
\includegraphics[scale=0.4]{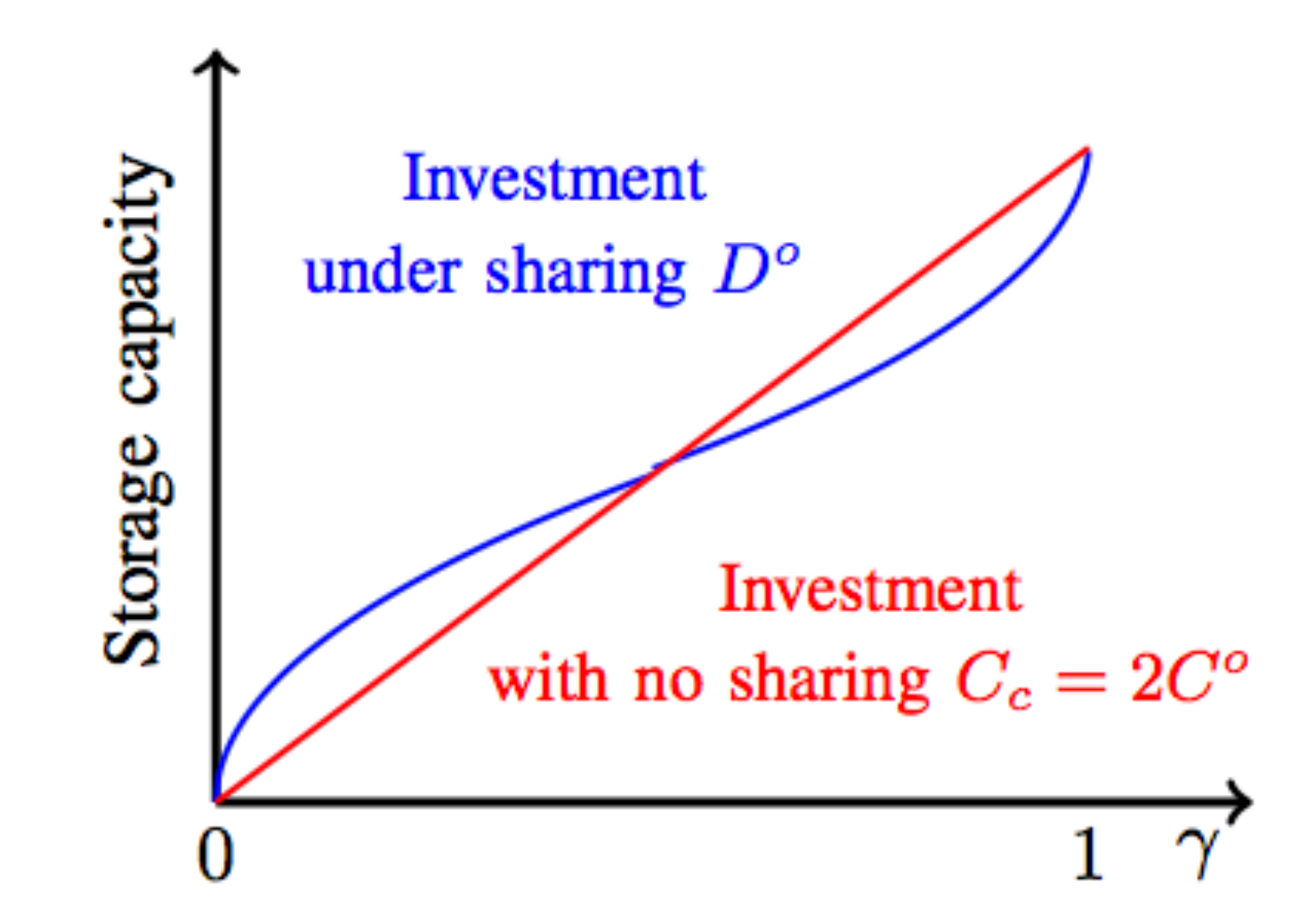}
\caption{Example: Under- and over-investment.} \label{fig:example}
\end{figure}

\begin{example}  \label{ex:1}
Consider two firms, indexed by $k=1,2$, whose peak period demands are the  random variables $X_1, X_2$ respectively.
Suppose $X_1, X_2$ are independent and uniformly distributed on $[0,1]$
Then, we have
\[  F_k(x) = \left\{ \begin{array}{cl} 0 & \text{ if } x < 0 \\ x & \text{ if } x \in [0, 1) \\ 1 & \text{ if } x > 1 \end{array} \right. \]
Fix $\gamma \in [0,1]$. The optimal storage investment of both firms is identical, and using Theorem \ref{thm:1}, we calculate this
to be  $C^o = F_k^{-1}(\gamma) =  \gamma$.
Their combined storage investment is 
\[ C_c = 2 C^o = 2 \gamma. \]
Now consider the entity formed by merging these firms. Sharing electricity between firms is an internal exchange within the entity.
This entity has combined peak period demand $X = X_1 + X_2$.
Notice that
\[ F_X(x) = \left\{ \begin{array}{cl}
0 & \text{ if } x<0  \\ 0.5 x^2 & \text{ if } x \in [0,1) \\ 1 - 0.5(x-2)^2 & \text{ if } x \in [1,2) \\ 1 & \text{ if } x \geq 2
\end{array} \right. \]
The optimal storage investment of the aggregate entity is $D^o = F_X^{-1}(\gamma)$ which is
\[ D^o = F_X^{-1}(\gamma) =  \left\{ \begin{array}{cl}
\sqrt{2 \gamma} & \text{ if } \gamma \in [0, 0.5]   \\ 2 + \sqrt{2 - 2\gamma} & \text{ if } \gamma \in [0.5,1]   
\end{array} \right. \]
Plotted in Figure \ref{fig:example} are $C_c$, and $D^o$ as functions of $\gamma$. Notice that $D^o < C_c$ when $\gamma < 0.5$ and 
$D^o > C_c$ if $\gamma > 0.5$. \hfill $\Box$
\end{example}


\begin{remark}
The example above reveals that without sharing, 
firms might over-invest in storage because they are going it alone and do not have the opportunity to
buy stored electricity from other firms. This happens when $\gamma$ is large.
They might also under-invest because they forgo revenue opportunities that arise from selling their stored electricity to other firms.
This happens when $\gamma$ is small. \hfill $\Box$
\end{remark}


\section{Main Results: With Sharing} \label{sec:noshare}

Consider  $n$ firms. Firm $k$ has chosen to invest in $C_k$ units of storage to arbitrage against the ToU pricing it faces.
On a given day, suppose the total peak-period energy demand of firm  $k$ is $X_k$.
The firm will choose to first service $X_k$ using its cheaper stored energy. This may leave a surplus of stored energy $(C_k - X_k)^+$.
This excess energy available to firm $k$ in its storage can be sold to other firms. Conversely, it may happen that firm $k$ faces a deficit in its demand
of $(X_k - C_k)^+$
even after using its stored energy. This deficit could be purchased from other firms that have a surplus, or from the utility.

\subsection{The Spot Market for Stored Energy}

We consider a spot market for trading excess energy in the electricity storage of the collective of firms.
Let $S$ be the total supply of energy available from storage 
from the collective after they service their own peak period demand. Let $D$ be the total deficit of energy that must be acquired 
by the collective of firms after they service their own peak period demand. So,
\[ S = \sum_k (C_k - X_k)^+, \quad D = \sum_k (X_k - C_k)^+ \]

If $S > D$, the suppliers compete against each other and drive the price down
to their (common) acquisition cost of $\pi_\ell$. Note that unsold supply is simply held. 
Since the storage is perfectly efficient and lossless (see Assumption A.4), there are no holding costs.
This is equivalent to selling unsold supply at $\pi_\ell$ to an imaginary firm, and buying it back during the next off-peak period at price $\pi_\ell$.
As a result,  storage is {\sl completely discharged} during the peak period, and {\sl fully recharged} during the subsequent off-peak period.
The entire supply $S$ is traded at $\pi_\ell$ if $S>D$.

If $S < D$, some electricity must be purchased from the utility which is the supplier of last resort. 
Consumers compete and drive up the price $\pi_h$  offered by the utility. As a result, the excess energy $S$ in storage is traded at
$\pi_h$ when $S<D$.

\begin{figure}[h!]
\captionsetup{justification=centering}
\centering
\includegraphics[scale=0.37]{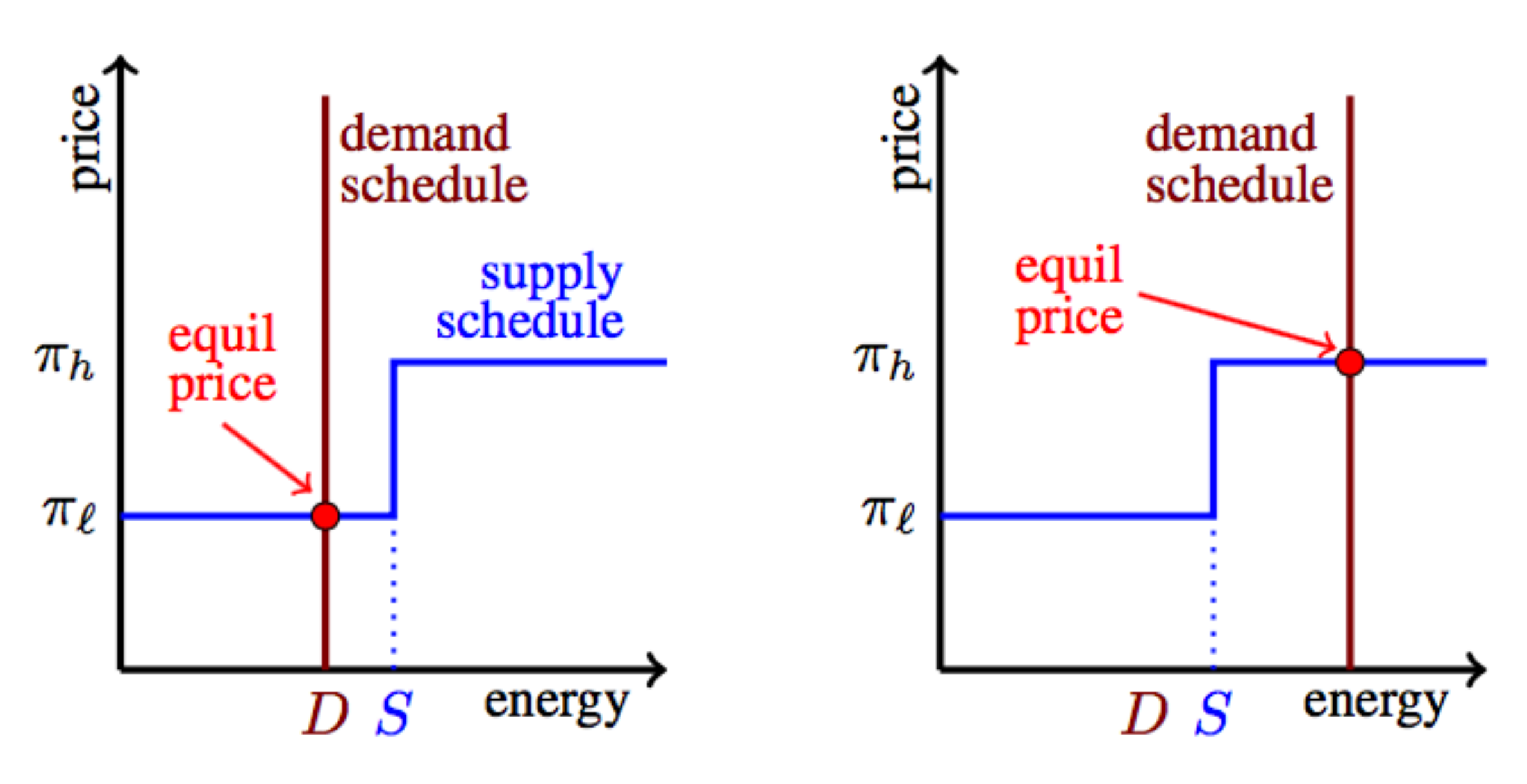}
\caption{Equilibrium Price: (a) left panel $S>D$, (b) right panel $S < D$. } \label{fig:equil}
\end{figure} 

The market clearing price is therefore
\beq \pi_{eq} = \left\{ \begin{array}{cl} \pi_\ell & \text{ if } S > D \\ \pi_h & \text{ if } S < D  \end{array} \right. \eeq
Note that the clearing price $\pi_{eq}$ is random and depends on supply-demand conditions in each peak period.
This competitive equilibrium is illustrated in Figure \ref{fig:equil}.
Note that
\[ 
\begin{aligned}
& & S - D & = \sum_k  \left( (C_k - X_k)^+   - (X_k - C_k)^+ \right) \\
& & & = \sum_k C_k - \sum_k X_k = C_c - X_c
\end{aligned} \]
where $C_c$ is the collective storage installed by the firms, and $X_c$ is their collective demand.
We can then re-write the clearing price as
\beq \label{eq:price}
\pi_{eq} = \left\{ \begin{array}{cl} \pi_\ell & \text{ if } C_c > X_c \\ \pi_h & \text{ if } C_c < X_c  \end{array} \right. \eeq

\subsection{Optimal Investment Decisions under Sharing}

Consider a collection on $n$ firms. Suppose firm $k$ has chosen to invest in $C_k$ units of storage.
The expected daily cost for firm $k$ under sharing is
\beq  \label{eq:costb}
J_k(C_k, C_{-k}) =  \underbrace{\pi_s C_k}_\text{\footnotesize cap ex} + \underbrace{\pi_\ell C_k}_\text{\footnotesize recharge} 
 + \  \underbrace{\EXP{\pi_{eq} (X_k - C_k) }}_\text{\footnotesize trade surplus/deficit}
\eeq
Note that this expected cost depends on the decisions $C_{-k}$ of all the other firms. This dependence appears implicitly through the
random clearing price $\pi_{eq}$ for shared energy. 

Suppose firms $i, i\neq k$ have invested in $C_i$ units of storage. 
The optimal investment of firm $k$ under  sharing is to purchase $C_k^o$ units of storage where
\[ C_k^o = \arg\min_{C_k} J(C_k, C_{-k}) \]
Unfortunately, the cost function $J_k$ is, in general, non-convex. It may have multiple local minima, and nonunique global minimizers.
Determining $C_k^o$ analytically can be extremely difficult. Remarkably, in a non-cooperative game setting when all firms seek to miminize their cost, we can explicitly characterize optimal investment decisions (see Theorem \ref{thm:uniqueness}).

\subsection{The Social Cost}

The {\sl social cost} is the sum of the daily expected costs of all the firms:
\begin{eqnarray} \nonumber
J_c(C_1, \cdots, C_n) & = & \sum_k J(C_k, C_{-k}) \\ \nonumber
 & = & \pi_\ell C_c + \pi_\ell \EXP{Y_c} + \EXP{\pi_{eq} (X_c - C_c) } 
\end{eqnarray}
We can view trading excess storage as internal transactions within the collective.
From this observation, it is straightforward to verify that the social cost depends only on the collective
investment $C_c$ and that
\begin{eqnarray} \nonumber \label{eq:costc}
 J_c(C_c) & = &  \pi_s C_c + \pi_\ell \EXP{Y_c} + \pi_h \EXP{(X_c - C_c)^+}  \\ 
 & & \quad + \  \pi_\ell \EXP{\min\{C_c, X_c\}}
\end{eqnarray}
This can be regarded as the daily expected cost of the collective firm under no sharing (see  (\ref{eq:costa})). 
A social planner would minimize the social cost and select $C_c = C^*$ where $F_c(C^*) = \gamma$ (see Theorem \ref{thm:1}).
No prescription would be made on allocating the total investment $C^*$ among the firms.

\subsection{Non-cooperative Game Formulation}

We stress that the optimal storage decision $C_k^o$ of firm $k$ depends on the investment choices $C_i, i\neq k$ made by all other firms.
This leads naturally to a non-cooperative game-theoretic formulation of the {\em Storage Investment Game} $\Gcal$. The players are the $n$ firms.
The decision of firm $k$ is $C_k$ and its cost function is $J_k(C_k, C_{-k})$.  We explore pure strategy Nash equilibria for this game.

\begin{theorem}\label{thm:existence}
Suppose for $k=1,\cdots, n$,
\beq \label{eq:condn}
\frac{d \EXP{ X_k  \mid X_{c} = \beta}}{d \beta} \geq 0
\eeq
Then the Storage Investment Game admits a Nash Equilibrium.
\end{theorem}

\begin{remark}
The {\em alignment condition} (\ref{eq:condn}) needed to establish existence of a Nash equilibrium has a natural interpretation - 
the expected demand $X_k$ of firm $k$ increases if the total demand $X_c$ increases. 
This is not unreasonable. This is not unreasonable. For example, if $X_{k}$ and $X_{c}$ are jointly Gaussian, then (\ref{eq:condn}) hold if they are positively correlated.   \hfill $\Box$
\end{remark}

\begin{figure}
\captionsetup{justification=centering}
\centering
\includegraphics[scale=0.65]{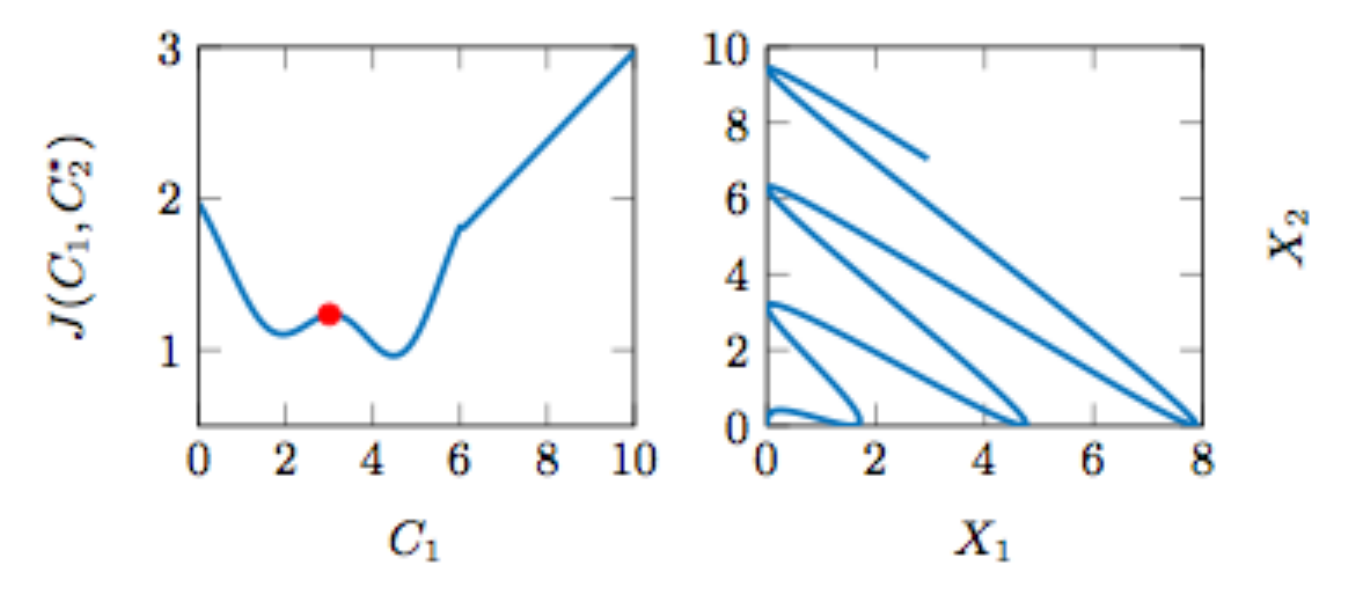}
\caption{Example: no Nash Equilibrium. } \label{fig:nonash} 
\end{figure}

\begin{example}
We can construct examples with exotic demand distributions where a Nash equilibrium does not exist. 
 Let $W$ be a random variable uniformly distributed on $[0,10]$.
Define the peak period consumption for two firms by
\[
X_1 = W \sin^2(W), \ \ X_2 = W \cos^2(W)
\]
Notice that the collective demand is $X_c = X_1 + X_2 = W$. The support of $X_1,X_2$ in the plane in shown in the right panel of Figure \ref{fig:nonash}.
We calculate
\[ \EXP{X_1 \mid X_1 + X_2 = \beta} = \beta \sin^2(\beta) \]
This is not a non-decreasing function of $\beta$. Thus the alignment condition (\ref{eq:condn}) is violated.

We choose $\pi_s = 0.3, \pi_\delta = 1$. Then, $F_c(Q) = 0.7 = \beta$, which implies $Q = 7$.
If a Nash equilibrium exists, it is unique and must be given by (see Theorem \ref{thm:uniqueness})
\begin{eqnarray*}  C_1^* & = & \EXP{X_1 \mid X_c = 7} = 7\sin^2(7) = 3.02 \\  C_2^* & = &  \EXP{X_2 \mid X_c = 7} = 7 \cos^2(7) = 3.98
\end{eqnarray*}
The cost function for firm 1 is
\[ J_1(C_1, C_2^*) = 0.3 C_1 + 0.1 \int_{C_1+3.98}^{10} (W\sin^2(W) - C_1) dW \]
This function is plotted in the left panel of Figure \ref{fig:nonash}.  Notice that $C_1^*$ is a {\em not} a global minimizer,
proving that a Nash equilibrium does not exist.
\hfill $\Box$
\end{example}

%

\begin{theorem}\label{thm:uniqueness}
Suppose the Storage Investment Game admits a Nash Equilibrium. Then it is unique and given by
\beq \label{eq:share}
C^*_k = \mathbb{E}[X_k|X_c=Q],\ k=1,\cdots,n 
\eeq
where $Q$ is the unique solution of
\beq \label{eq:share2}
F_c(Q) = \frac{\pi_\delta - \pi_s}{\pi_\delta} = \gamma 
\eeq
The resulting optimal cost is
\beq \label{eq:optcostb} J^* = \pi_\ell \EXP{X_k} + \pi_s \EXP{X_k \mid X_c \geq C^*} \eeq
\end{theorem}

\begin{remark}
The game $\Gcal$ is {\em nonconvex}. It is remarkable that it admits an explicit  characterization of its unique Nash equilibrium should one exist.
It commonly happens that the cost function $J_k(C_k,C_{-k}^*)$ of firm $k$ given optimal decisions of other firms has multiple local minima. It is surprising that $C^*_k$ is the {\em unique global minimizer} of this function. \hfill $\Box$
\end{remark}



\begin{remark}
Our problem formulation above {\em does not} assume a perfect competition model. 
Indeed, under perfect competition, the celebrated welfare theorems 
assure existence of a unique Nash equilibrium. In our analysis, we allow firm $k$ to take into account the influence its investment
decision $C_k$ has on the statistics of the clearing price $\pi_{eq}$. 
This is a Cournot model of competition \cite{Tirole}, under which Nash equilibria do not necessarily exist. \hfill $\Box$
\end{remark}


\begin{theorem}\label{thm:property}
The unique Nash equilibrium of Theorem \ref{thm:uniqueness} has the following properties:
\balphlist
\item The Nash equilibrium supports the social welfare: the collective investment $C_c$ of the firms coincides 
with the optimal investment of the collective firm with peak period demand $X_c = \sum_k X_k$. 
\item Individual rationality: No firm is better off on its own as a standalone firm without sharing.
\item Coalitional stability: Assume the alignmnet condition (\ref{eq:condn}) holds. No subset of firms is better off defecting to form their own coalition. 
\item No arbitrage: At the  Nash equilibrium, we have
$\dst{ \EXP{\pi_{eq}} = \pi_s + \pi_\ell}$.
\item Neutrality of Aggregator: If firm $k$ has  peak-period consumption $X_k = 0$, it will not invest in storage, i.e. $C_k^* = 0$.
\end{list}
\end{theorem}

\begin{remark} 
Part (e) of this Theorem establishes that there is no pure-storage play. A firm that does not consume electricity in the peak period
has no profit incentive to invest in storage. The aggregator is in this position. 
An important consequence is that the aggregator is in a position of neutrality with respect to the firms.
It can therefore act to supply the information necessary for firm $k$ to make its optimal investment choice.
This information consists of (i) the joint statistics of $X_k$ and $X^c$, and (ii) the cumulative optimal investment $C^c$ of all the firms.
With this information, firm $k$ can compute its share of the optimal storage investment $C^c$ as in (\ref{eq:share}). As a result,
the private information $X_i, i\neq k$ of the other firms is protected. The neutrality of the aggregator affords it a position 
to operate the market, determine the market clearing price of shared storage, conduct audits, and settle transactions.
\hfill $\Box$
\end{remark}

\begin{remark}
We can also treat storage sharing in a cooperative game formulation. In this setting, we infer from Theorem \ref{thm:property} that
the allocation $a_k = -J_k(C_k^*,C_{-k}^*)$ to firm $k$ is an imputation. It is budget balanced, individually rational, and coalitionally stable.
This imputation therefore lies in the core of the cooperative game. There may be other imputations in the core. \hfill $\Box$
\end{remark}

\section{Non-ideal Storage} \label{sec:gen}

We generalize our results to accommodate certain aspects of non-ideal storage. 
Let  $\eta_i, \eta_o$ be the charging and discharging efficiency respectively. 
We do not address maximum rates of charge or discharge, or treat leakage.
Incorporating leakage is challenging because it affects the control strategy of storage and the clearing price in the spot-market.

\begin{theorem} \label{thm:lossy}
With non-ideal storage, the optimal decision of a firm under no sharing is to invest in $C^o$ units of storage where
\beq F(\eta_o C^o) = \frac{\pi_h\eta_o - \pi_\ell /\eta_i -\pi_s}{\pi_h\eta_o - \pi_\ell /\eta_i} \eeq
\end{theorem}

Charging inefficiency has the effect of inflating the off-peak price $\pi_\ell$, and discharging inefficiency discounts the peak-period price $\pi_h$.
These together reduce the arbitrage opportunity. Our results on optimal decisions under sharing also generalize easily.

\section{Joining the Club} \label{sec:club}

We have thus far assumed that all firms make their storage investment decisions simultaneously.
A better model would allow for sequential capital investment decisions. 
We explore the situation when a collective of firms $\Ccal$ has made optimal storage investments, and a new firm $F_{n+1}$ wishes to join.

\begin{theorem} \label{thm:extensive}
Consider a collective of $n$ firms. Let $Q^n$ be the optimal combined storage investment of these firms under sharing. 
Suppose a new firm joins the collective. Let $Q^{n+1}$ be new combined optimal storage investment of these $n+1$ firms
under sharing. Then,
\[ Q^{n+1} \geq Q^n. \] 
\end{theorem}

This result shows that the optimal storage investment is {\sl extensive}. As firms join the collective, the optimal storage investment 
must increase. As a result, the collective of firms does not have to divest any storage investments already made. It must merely purchase
$Q^{n+1} - Q^n$ additional units of storage. 
In addition, the optimal storage investments of firms in $\Ccal$ change when the new firm joins the collective.
As a result, these firms will have to rearrange their fraction of storage ownership requiring an internal exchange of payments.
If the storage is co-located and managed at the aggregator, this rearrangement reduces to simple financial transactions.

It is clear from the coalitional stability result of Theorem \ref{thm:property}(b), that both the original collective $\Ccal$ 
and the new firm are better off joining forces.
However, simple examples reveal that some individual firms in $\Ccal$ may be worse off in the expanded collective $\Ccal \cup F_{n+1}$.
This raises interesting issues on voting rights. Under veto power the new firm may not be invited to expand the coalition. Under
a cost-weighted majority vote, the new firm will always be invited to expand the coalition. These questions require further exploration.

\section{Simulation Studies}
\label{simulation}

\begin{figure}
\captionsetup{justification=centering}
\centering
\includegraphics[scale=0.45]{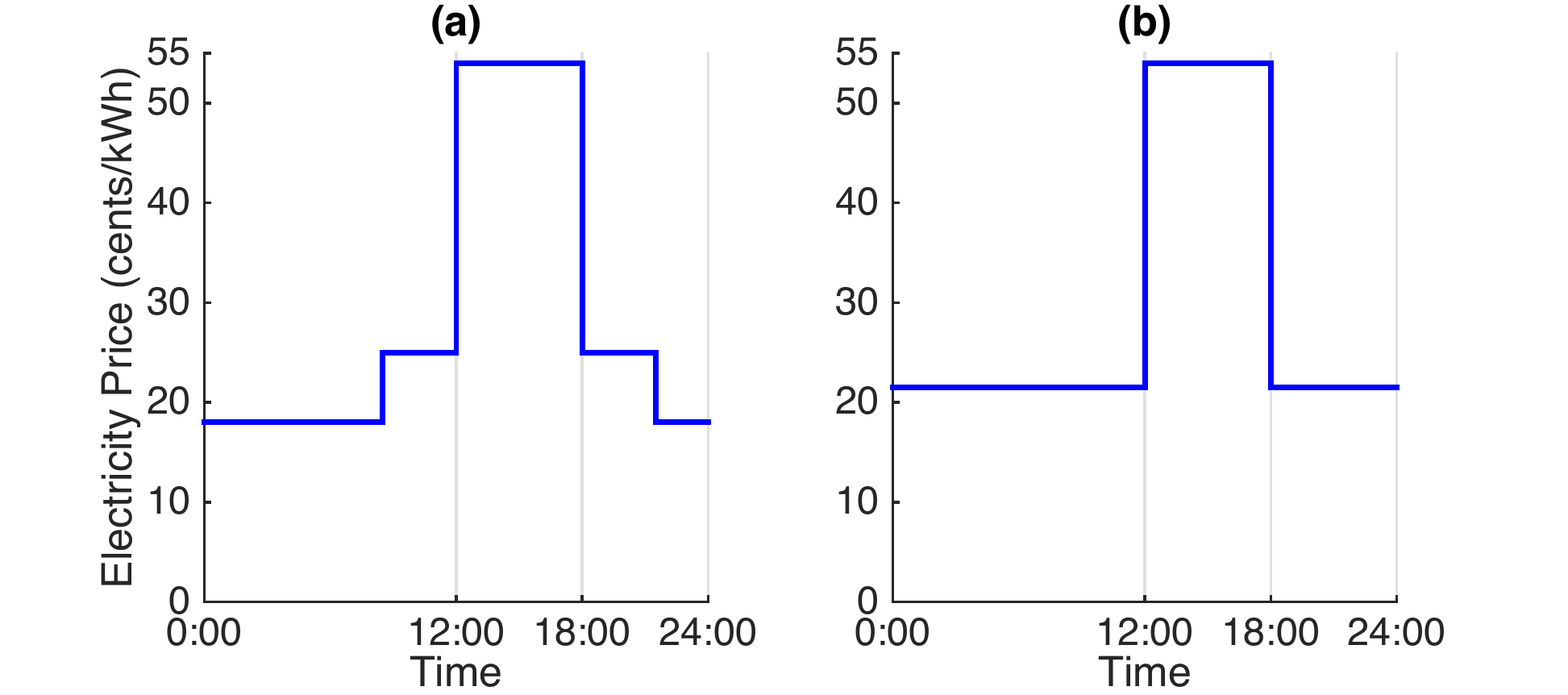}
\caption{ToU pricing: (a) real three-period pricing, (b) simplified two-period pricing.} \label{fig:real_tou}
\end{figure}

\begin{figure}
\captionsetup{justification=centering}
\centering
\includegraphics[scale=0.5]{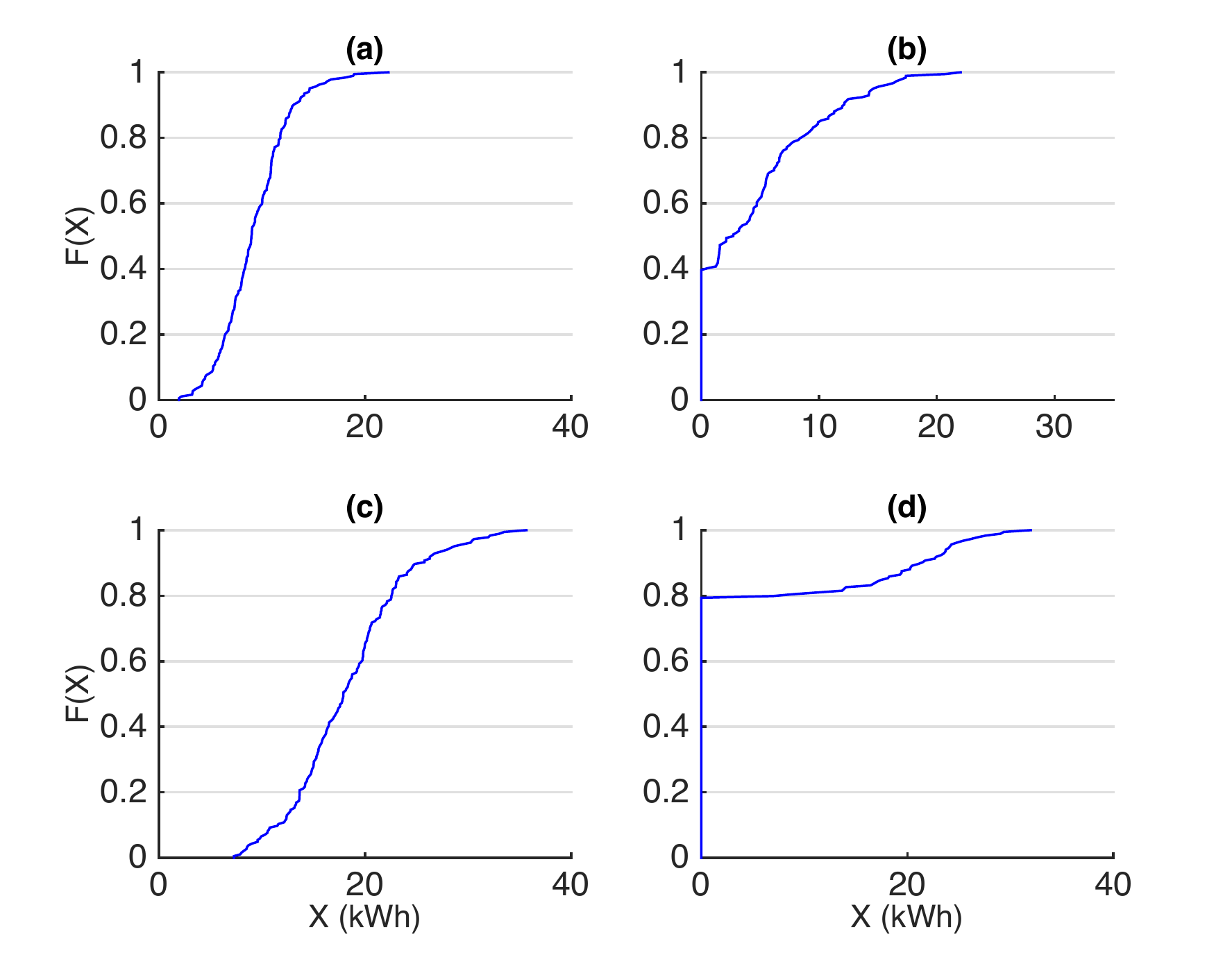}
\caption{Sample cumulative distribution functions of peak-period demand $X_k$ for 4 households.} \label{fig:sample}
\end{figure}


We illuminate the analytical development of sharing storage using synthetic simulations.
Figure \ref{fig:real_tou}(a) shows the three-period  (peak, partial peak, and off-peak) A6 tariff offered by PG\&E during summer months. 
We approximate this by the two-period ToU tariff shown in Figure \ref{fig:real_tou}(b). 
We set $\pi_h=54$\textcent/kWh, and $\pi_\ell=21.5$\textcent/kWh (average of the partial peak and off-peak prices). 
We use the publicly available Pecan Street data set \cite{pecan} which offers 1-minute resolution consumption data from 1000 households.
From historical data, we use  standard methods to estimate demand statistics.
Figure \ref{fig:sample} shows sample empirical cumulative distribution functions (cdfs) $F_k(\cdot)$ for 4 household $k=1, \cdots, 4$.
It is apparent that there is considerable statistical diversity in peak-period consumption. 
Panel (a) shows a representative cdf,  panel (b) suggests that some households are consistently vacant during peak hours, panel (c) shows a 
constant background load, panel (d) suggests some users have low variability in  their peak-period demand.

\begin{figure*}[!htb]
\minipage{0.32\textwidth}
  \includegraphics[width=\linewidth]{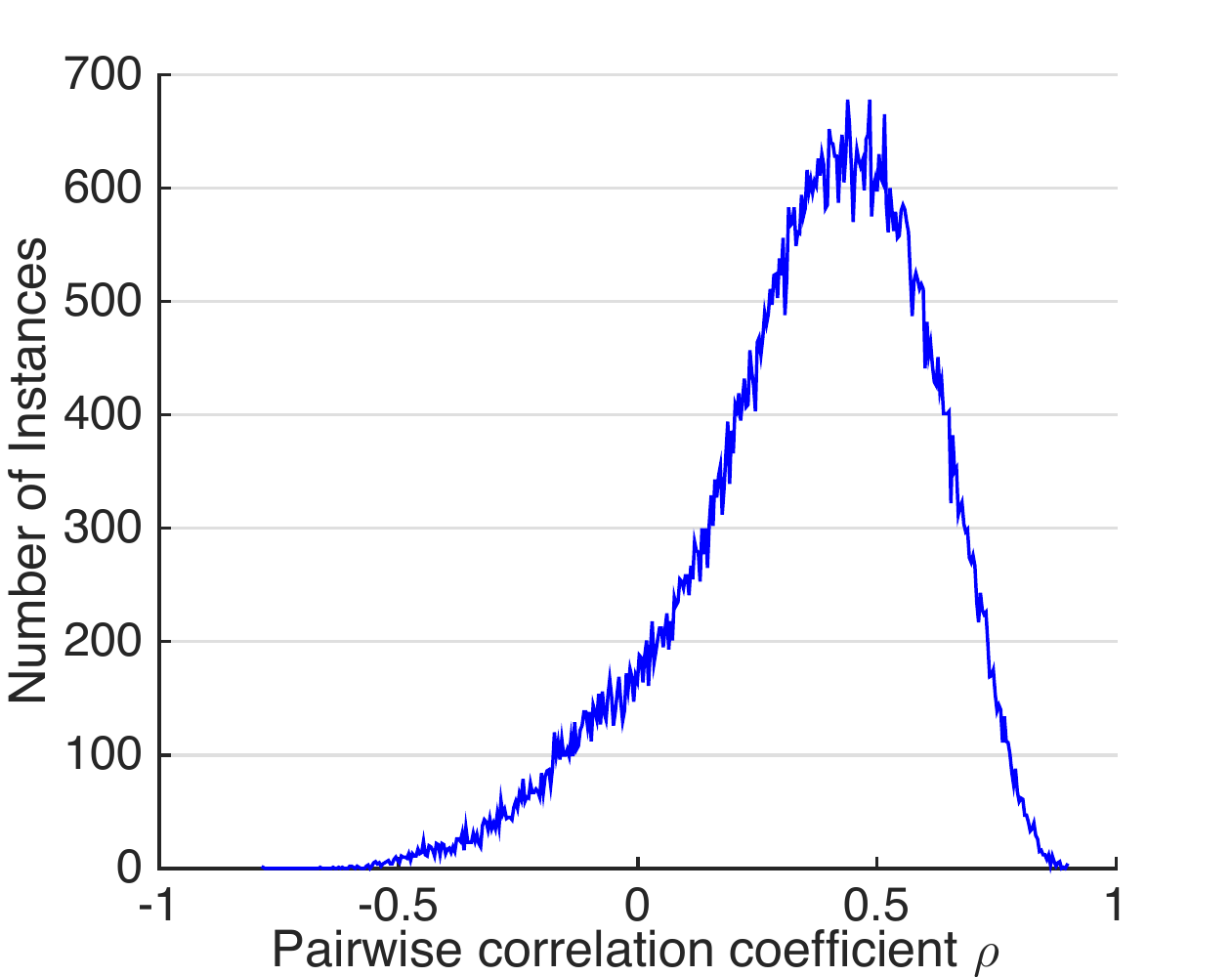}
  \caption{Histogram of pairwise correlation coefficients.}\label{fig:PecanData_correlation}
\endminipage\hfill
\minipage{0.32\textwidth}
  \includegraphics[width=\linewidth]{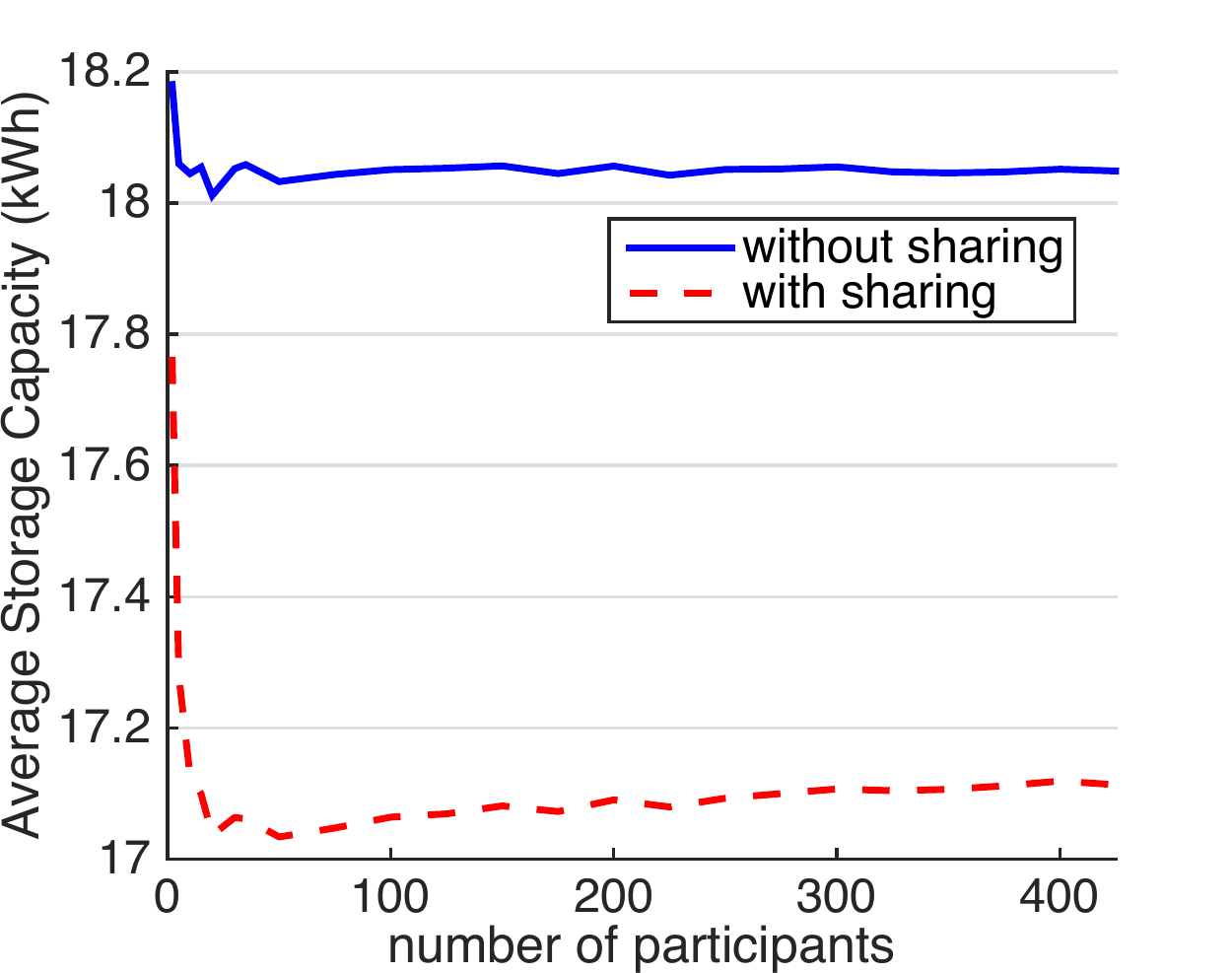}
  \caption{Average optimal investment.}\label{fig:aveStorage}
\endminipage\hfill
\minipage{0.32\textwidth}%
  \includegraphics[width=\linewidth]{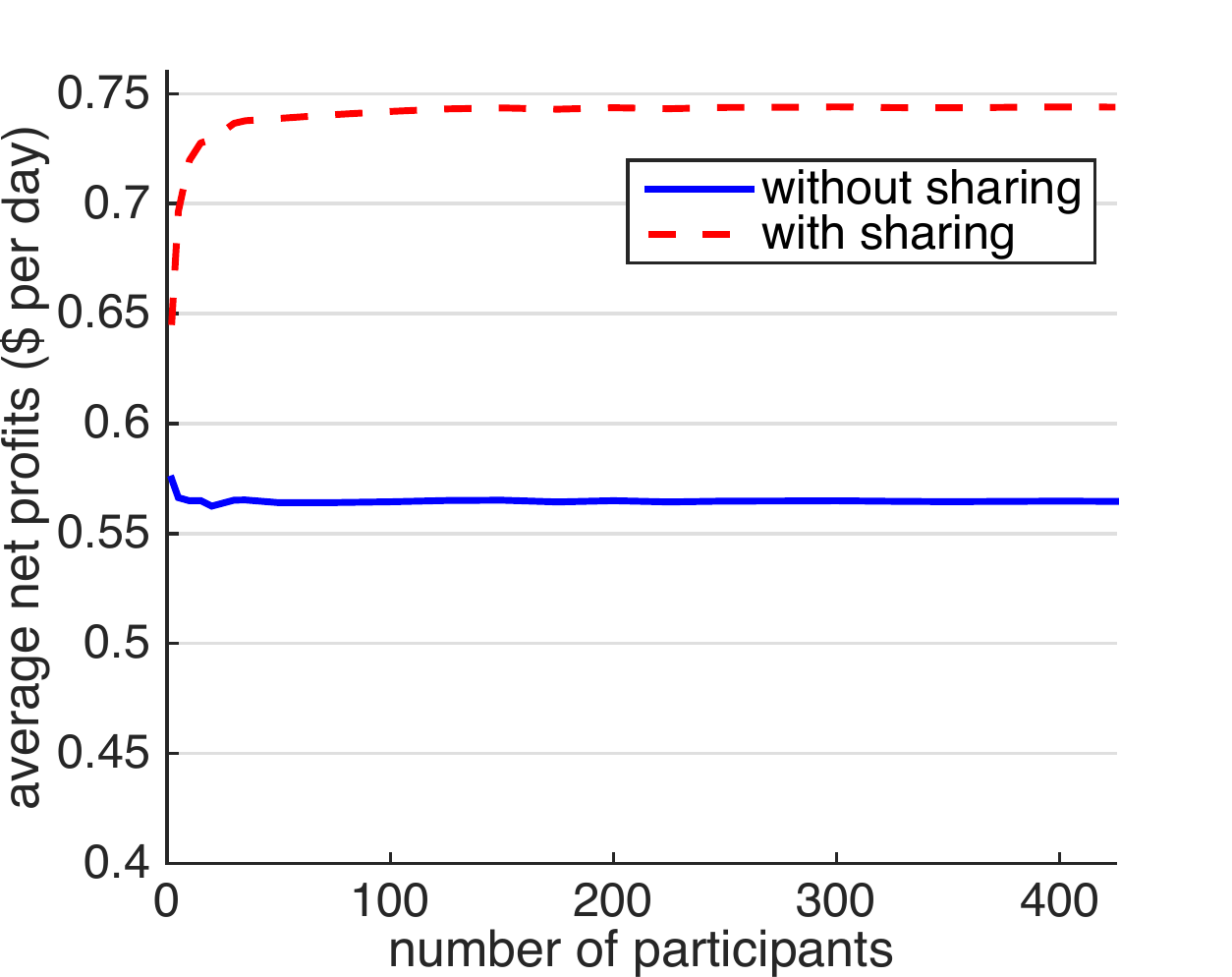}
  \caption{Average savings using storage for ToU price arbitrage.}\label{fig:aveSavings}
\endminipage
\end{figure*}

For storage sharing to be beneficial, we require statistical diversity in the peak period demands $X_k$.  
A histogram of the pairwise correlation coefficients is shown in Figure \ref{fig:PecanData_correlation}. 
The average pairwise correlation coefficient between households is approximately $0.5$, and there are many pairs of households with negatively correlated demands. 

%
We compare two cases: (a) {\sl without sharing}, and (b) {\sl with sharing}.
We have already seen (see Example \ref{ex:1}) that firms may over- or under-invest under no sharing depending on the statistics of their demand.
For this data set, Figure \ref{fig:aveStorage} shows that the average storage investment is $\approx$5\% lower under sharing.
Finally, we compute the average financial benefit under sharing. The expected daily cost of a household  that chooses not to invest in storage is 
$J = \pi_h \EXP{X_k}$. If this household invests optimally in storage, but does not participate in sharing, its expected daily cost is (see equation (\ref{eq:optcosta}))
\[ J^o = \pi_\ell \EXP{X_k} + \pi_s \EXP{X_k \mid X_k \geq C^o} \]
where $C^o$ is prescribed by Theorem \ref{thm:1}. If this household invests optimally while participating in the spot market for storage, its expected daily cost is (see equation (\ref{eq:optcostb}))
\[ J^* = \pi_\ell \EXP{X_k} + \pi_s \EXP{X_k \mid X_c \geq C^*} \]
as shown in Theorem \ref{thm:uniqueness}.
The expected daily arbitrage revenue from using storage without sharing is $\Delta^{ns} = J - J^o$, and is
$\Delta^s  = J - J^*$ under sharing. 
We plot $\Delta^{ns}, \Delta^s$ averaged across users. 
Figure \ref{fig:aveSavings} shows that although the users earn $\approx$55\textcent \ per day without sharing on average vs $\approx$75\textcent \ under sharing
which represents a 50\% better return that without sharing. 

%
%
%

\section{Physical and Market Implementations}

Sharing electricity services requires coordination. For example, if the service is a delivery of power from one rooftop PV to a remote consumer in the community, signals must be sent to coordinate the equipment of both. Set point schedules for inverters must be communicated in advance of the physical exchange of energy.

Our focus has been on sharing {\em energy} services. The transactions involved are exchanges of energy during the peak period 
without stipulating {\em when} this energy should be delivered. Small businesses could invest in electricity storage for many reasons 
beyond time-of-use price arbitrage, including smoothing high-frequency variations in PV production, or protection from critical peak pricing (CPP). 
CPP \cite{CPP} is a common tariff where firms face a substantial surcharge based on their monthly peak demand. 
Sharing surplus energy in electricity storage does not preclude these other applications.

In our framework, electricity storage could be physically distributed among the firms, or centralized.  
A distributed storage architecture requires $n$ inverters and results in larger losses.
Centralized storage co-located, installed, and managed by the aggregator, requires a single inverter and promises economies of scale. 
Firms can lease their fair share of storage capacity directly from the aggregator.

We have studied a spot market for trading energy in electricity storage. Other arrangements are possible -- bilateral trades, 
auctions, or bulletin boards to match buyers and sellers. In any event, realizing a sharing economy for electricity services requires
a scalable software platform to accept supply/demand bids, clear markets, publish prices, and conduct audits.

The physical and market infrastructure necessary to support the broader sharing economy for the smart grid is a topic that demand deeper exploration.

\bibliographystyle{IEEEtran}


\section*{Appendix:Proofs}

\subsection{Proof of Theorem \ref{thm:1}}

The daily expected cost  for a single firm under no sharing is (see equation (\ref{eq:costa}))
\begin{eqnarray*} 
J(C) & = & \pi_s C + \EXP{\pi_h (X - C)^+ + \pi_\ell Y + \pi_\ell \min\{C, X\}} \\
& = & \pi_s C + \pi_h \int_C^\infty (x-C) f_X(x) dx  + \EXP{\pi_\ell Y} \\
& & \quad + \pi_\ell C \cdot \Prob{X \geq C} + \pi_\ell \int_0^C x f_X(x) dx
\end{eqnarray*}
It is straightforward to verify that this is strictly convex in $C$. As a result, the optimal investment $C^o$ is the unique solution of
the first-order optimality condition
\begin{eqnarray*}
0 & = & \frac{d J}{dC} = \pi_s - \pi_h \int_C^\infty  f_X(x) dx + \pi_\ell \Prob{X \geq C}  \\
& = & \pi_s + (\pi_\ell -  \pi_h) \left(1 - F(C)\right)
\end{eqnarray*}
Rearranging this expression yields
\[ F(C) = \frac{\pi_s - \pi_\delta}{\pi_\delta} \]
proving the claim. \hfill $\Box$

\subsection{Proof of Theorem \ref{thm:existence}} 

Consider firm $k$. Its decision is $C_k$. Fix the decisions of all other firms $C_{-k}^*$ where
\[ C_i^* = \EXP{X_i \mid X_c = Q}, \ \ F_c(Q) = \gamma = \frac{\pi_\delta - \pi_s}{\pi_\delta} \]
Simple algebra reveals that 
\[ \pi_s - \pi_\delta + \pi_\delta F_c(Q) = 0 \]
Define the quantity
\[ \alpha = \sum_{i \neq k} C_i^* \]
The expected daily cost of firm $k$ defined on $C_k \geq 0$ is 
\begin{eqnarray*}
&& J_k(C_k \mid C_{-k}^*) =  \pi_s C_k  + \pi_\ell C_k
 + \EXP{ \pi_{eq}(X_k - C_k)}
 \end{eqnarray*}
We have to show that $C^*_k$  is a global minimizer of $J_k(C_k, C_{-k}^*)$.

Using the characterization (\ref{eq:price}) of the spot market clearing price $\pi_{eq}$, 
and noting that the statistics of $X_k$ are not influenced by $C_k$ (see Assumption A.3), this cost function simplifies to: 
\[ 
\pi_s C_k   + \  \pi_\delta  \mkern-18mu \int\displaylimits_{X_k = 0}^\infty \int\displaylimits_{\quad X_c = C_k + \alpha}^\infty  \mkern-18mu 
\left(X_k - C_k\right) f_{X_k X_c} (X_k, X_c) dX_k dX_c  
\]
It is easy to show using the Leibniz rule that
\begin{eqnarray} \nonumber
\frac{d J_k}{dC_k} & = &  -  \ \pi_{\delta} \cdot  f_{c}(C_k + \alpha) \cdot 
\underbrace{\left(\EXP{ X_k \mid X_{c} = C_k + \alpha} - C_k \right)}_{\psi(C_k)}  \\ 
& & 
+ \ \underbrace{\pi_{s} - \pi_{\delta} + \pi_\delta F_c(C_k + \alpha)}_{\phi(C_k)} 
\end{eqnarray}
We explore properties of the functions $\phi$ and $\psi$. 
We have
\begin{eqnarray} \label{eqx}
\phi(C_k) & & \text{ is monotone increasing in } C_k \\  \nonumber
\phi(C^*_k) & = & \pi_{s} - \pi_{\delta} + \pi_\delta F_c(C^*_k + \alpha) \\ \label{eqy}
& = & \pi_{s} - \pi_{\delta} + \pi_\delta F_c(Q) = 0 
\end{eqnarray}
Here, we have made use of the assumption that $f_c(\cdot) > 0 $. Thus $\phi$ is monotone increasing and vanishes at $C_k = C^*_k$.
Next observe that
\begin{eqnarray*}
 \beta & = & \sum_k \EXP{ X_k  \mid X_{c} = \beta} \\
 1 & = &  \sum_k \underbrace{ \frac{d \EXP{ X_k  \mid X_{c} = \beta}}{d \beta} }_{\geq 0 \text{ by equation \ref{eq:condn}}}
  \implies  \frac{d \EXP{ X_k  \mid X_{c} = \beta}}{d \beta} \leq 1
 \end{eqnarray*}
Here we have made critical use of the technical condition (\ref{eq:condn}) needed to establish existence of a Nash equilibrium. It then follows that
\begin{eqnarray*}
\frac{ d\psi(C_k)}{dC_k} & = & \frac{d \EXP{ X_k  \mid X_{c} = \beta}}{d \beta} - 1 \leq 0 \\ 
\psi(C^*_k) & = & \EXP{ X_k  \mid X_{c} = C^*_k + \alpha} - C^*_k \\ 
& = & \EXP{ X_k  \mid X_{c} = Q} - C^*_k  = 0
\end{eqnarray*}
Thus $\psi$ is monotone non-increasing and vanishes at $C_k = C^*_k$.
As a result, 
\[ - \pi_{\delta} \cdot  f_{c}(C_k + \alpha) \cdot \psi(C_k) \left\{ \begin{array}{cl} \leq 0 &  C_k \leq C^*_k  \\ \geq0  & C_k \geq C^*_k  \end{array} \right. \]
Combining this with the properties of $\phi$ in equations (\ref{eqx}, \ref{eqy}), we get
\[
\frac{d J_k}{dC_k} 
\left\{ \begin{array}{cl} < 0 &  C_k < C^*_k  \\  = 0 & C_k = C^*_k \\ >  0  & C_k > C^*_k  \end{array} \right. \]
This proves that $C^*_k$ is the global minimizer of $J_k(C_k, C^*_{-k})$, establishing that $C^*$ is a Nash equilibrium. \hfill $\Box$

\subsection{Proof of Theorem \ref{thm:uniqueness}}  

Let $D_k, k = 1,\cdots, n$ be any Nash equilibrium. We show that $D = C^*$ where
\[ C_k^* = \EXP{X_k \mid X_c = Q}, \ \ F_c(Q) = \gamma = \frac{\pi_\delta - \pi_s}{\pi_\delta} \]
Simple algebra reveals that $Q$ is the unique solution of 
\[ \pi_s - \pi_\delta + \pi_\delta F_c(Q) = 0 \]

Let $\beta = \sum_k D_k$, and define the constants
\begin{eqnarray*} 
K_1  & = &  \pi_s - \pi_\delta + \pi_\delta F_c(\beta) \\
K_2 & = & \pi_{\delta}   f_{X_{c}}(\beta) > 0
\end{eqnarray*}
Define the index sets
\[ \Mset = \{i: D_i > 0\}, \quad \Nset = \{j: D_j = 0\} \]
Since $D$ is a Nash equilibrium, it follows that  $D_k$ is a global minimizer of $J_k(C_k \mid D_{-k})$.
We write the first-order optimality conditions for $i \in \Mset $:
\begin{eqnarray}  \nonumber
0 & = & \left.\frac{d J_i(C_i \mid D_{-i}) }{dC_i}\right|_{D} \\[0.1in] \label{eq:10a}
& = & K_1 - K_2 \cdot \EXP{ X_i - D_i \mid X_c = \beta}
\end{eqnarray}
The first-order optimality conditions for $j \in \Nset $ are:
\begin{eqnarray} \nonumber
0  & \leq & \left.\frac{d J_j(C_j \mid D_{-j}) }{dC_j}\right|_{D} \\[0.1in] \label{eq:10b}
& = & K_1  - K_2 \cdot \EXP{ X_j - D_j \mid X_c = \beta} 
\end{eqnarray}
Summing these conditions, we get
\begin{eqnarray*}
 0 & \leq &  n K_1 - K_2 \cdot \sum_{k=1}^n \EXP{ X_k - D_k \mid X_c = \beta} \\
  & = & n K_1 - K_2 \cdot \EXP{ X_c - \beta \mid X_c = \beta}  \\[0.1in]
  & = & n K_1 
\end{eqnarray*}
Thus, $K_1 \geq 0$. Using (\ref{eq:10a}), for any $i \in \Mset$ we write
\beq \label{eq:33}
\EXP{ X_i - D_i \mid X_c = \beta} = \frac{K_1}{K_2} \geq 0 
\eeq

Next, we have
\begin{eqnarray}   \nonumber
0 & = &  \sum_{k=1}^n \EXP{ X_k - D_k \mid X_c = \beta}  \\ \nonumber
& = &  \sum_{i \in \Mset} \EXP{ X_i - D_i \mid X_c = \beta}  +  \sum_{j \in \Nset} \EXP{ X_j \mid X_c = \beta} \\ \nonumber
& \geq & \sum_{i \in \Mset} \EXP{ X_i - D_i \mid X_c = \beta}  \\ \nonumber
& \geq & 0
\end{eqnarray}
Here, we have used (\ref{eq:33}) and the fact that the random variables $X_j$ are non-negative.
As a result, we have
\begin{eqnarray*}
 \text{for $i \in \Mset$:} & & \EXP{X_i - D_i \mid X_c = \beta} = 0 \\
\text{for $j \in \Nset$:} & & \EXP{X_j - D_j \mid X_c = \beta} =  \EXP{X_j \mid X_c = \beta} = 0
\end{eqnarray*}
So for $k = 1, \cdots, n$,
\[ 0 = \EXP{X_k - D_k \mid X_c = \beta} \iff D_k = \EXP{X_k \mid X_c = \beta} \] 
Using this in (\ref{eq:10a}) yields
\[ 0 = K_1 =  \pi_s - \pi_\delta + \pi_\delta F_c(\beta)  \]
This implies $\beta = Q$, and it follows that $D_k = C_k^*$ for all $k$, 
proving the claim. \hfill $\Box$

\subsection{Proof of Theorem \ref{thm:property}}

\balphlist
\item Notice that
\[ \sum_k C^*_k = \sum_k \EXP{X_k \mid X_c = Q} = \EXP{X_c \mid X_c = Q} = Q \]
Since $F_c(Q) = \gamma$, it follows that the Nash equilibrium (\ref{eq:share}) supports the social welfare.
\item We first show individual rationality - that no firm is better off defecting from the grand coalition. Consider firm $k$ on its own.
Its optimal investment decision is $C^o$ given by Theorem \ref{thm:1} and its optimal expected daily cost is $J^o$.
The standalone firm (a) buys its shortfall $(X_k - C_k)^+$ from the utility at $\pi_h$, and (b) spills its surplus $(C_k - X_k)^+$.
Under any sharing arrangement, firm $k$ benefits by (a) buying its shortfall at the possibly lower price $\pi_{eq}$, and (b) selling its surplus
at the possibly higher price $\pi_{eq}$. 
Suppose this firm were to retain its investment choice at $C^o$, but participate in sharing with the grand coalition. Its new cost function is
$J_k(C^o, C^*_{-k})$. Since any sharing arrangement reduces the cost of firm $k$, we have
\[ J^o \geq J_k(C^o,C^*_{-k}) \]
Since $C^*$ is a Nash equilibrium, we have
\[ J^o \geq J_k(C^o, C^*_{-k}) \geq J_k(C_k^*, C^*_{-k}) = J^*\]
As a result, $J^o \geq J^*$, proving the claim.

\item We next prove coalitional stability. 
Consider the Storage Investment Game $\Gcal$. We form coalitions $\Aset_j \subseteq \{1, \cdots, n\}$  such that
\[ \Aset_i \cup \Aset_j = \phi, \cup_k \Aset_k = \{1, \cdots, n \} \]
The game $\Gcal$ induces a new game $\Hcal$ with players $\Aset_i$ and associated cost
\[ J_{\Aset_i} = \sum_{k \in \Aset_i} J_k(C_1, \cdots, C_n) \]
Since the alignment condition (\ref{eq:condn}) holds for $\Gcal$, we have for any coalition $\Aset_i$, 
\[ \frac{d \EXP{X_{\Aset_i} \mid X_c = \beta}}{d \beta} = \sum_{k \in \Aset_i} \frac{d \EXP{X_k \mid X_c = \beta}}{d \beta} \geq 0 \]
Thus, the alignment condition holds for the induced game $\Hcal$. It therefore admits a unique Nash equilibrium $D^*$ where
\[ D^*_i = \EXP{X_{\Aset_i} |X_c = Q} = \sum_{k \in \Aset_i} C_k^* \]
Now individual rationality of $D^*$ in game $\Hcal$ is equivalent to coalitional stability of $C^*$ in game $\Hcal$, proving the claim.

\item Using the characterization (\ref{eq:price}) of $\pi_{eq}$, we have
\[ \EXP{\pi_{eq}} = \pi_\ell F_c(C_c) + \pi_h \left(1 - F_c(C_c) \right) = \pi_h - \pi_\delta F_c(C_c) \]
At the Nash equilibrium we have $C_c = Q$, where $F_c(Q) = \gamma = (\pi_\delta - \pi_s)/\pi_\delta$. Then,
\[ \EXP{ \pi_{eq} } = \pi_h - \pi_\delta + \pi_s  = \pi_\ell + \pi_s \]

\item Follows immediately from (\ref{eq:share}) with $X_k \equiv 0$.  \hfill $\Box$
\end{list}

\subsection{Proof of Theorem \ref{thm:lossy}}

The firm can withdraw at most $\eta_oC$ power from its fully charged storage. Therefore, the firm must purchase its peak-period deficit
$(X - \eta_oC)^+$ from the utility. During the off-peak period, the firm will fully recharge its storage because there are no holding costs (no leakage).
The cost function for the firm is then
\[
 J(C) = \pi_s C + \pi_h \EXP{(X - \eta_oC)^+} 
 + \  \frac{\pi_\ell}{\eta_i\eta_0}  \EXP{\min\{C, X\}}
\]
It is straightforward to verify that this function is strictly convex. Writing the first-order optimality condition, it follows that the optimal investment 
$C^o$ solves
\[ 
 0=\frac{dJ}{dC}=  \pi_s + \pi_h\eta_o \Pr{(X \geq \eta_oC)}  + \frac{\pi_\ell}{\eta_i}\Pr{(X \geq \eta_oC)}
\]
Rearranging terms establishes the claim. \hfill $\Box$


\subsection{Proof of Theorem \ref{thm:extensive}} 

First note that the storage investments $Q^n$ and $Q^{n+1}$ are optimal. Define the collective peak demands 
\[ A  = \sum_{k=1}^n X_k, \ \ B = A+ X_{n+1} \]
Using Theorem \ref{thm:1}, we have 
\[ F_{A}( Q^n) = F_{B}(Q^{n+1}) = \gamma \]
As a result,
\begin{eqnarray*}
\Prob{A \leq Q^n} & = &  \Prob{A + X^{n+1} \leq Q^{n+1}} \\
& \leq & \Prob{A \leq Q^{n+1}} 
\end{eqnarray*}
where the last inequality follows from $X_{n+1}\ge 0$ (demand is non-negative).
This forces
$Q^{n+1} \ge Q^n$, proving the claim.
\hfill $\Box$

%
%

\end{document}